\documentclass[a4paper, 10pt, oneside]{article}
\pdfoutput=1

\usepackage[sc]{mathpazo} 
\usepackage[T1]{fontenc} 
\usepackage{microtype} 

\usepackage[hmarginratio=1:1,top=32mm,columnsep=20pt]{geometry} 
\usepackage[hang, small,labelfont=bf,up,textfont=it,up]{caption} 
\usepackage{booktabs} 
\usepackage{float} 
\usepackage{hyperref} 

\usepackage{lettrine} 
\usepackage{paralist}

\usepackage{abstract} 

\usepackage{titlesec} 
\titleformat{\section}[block]{\large\scshape}{\thesection.}{1em}{} 
\titleformat{\subsection}[block]{\large}{\thesubsection.}{1em}{} 

\usepackage{fancyhdr} 
\usepackage{epsfig}

\usepackage{natbib}
\bibpunct{(}{)}{;}{a}{,}{,}
\usepackage{graphicx, color}

\usepackage{amsmath}

\title{\vspace{-15mm}\fontsize{24pt}{10pt}\selectfont
	\textbf{maigesPack: A Computational Environment for Microarray Data Analysis}
}

\author{
	\large
	\textsc{
		Gustavo H. Esteves\footnote{University of Para\'iba State, Campina Grande-PB, Brazil. email: 
			\href{mailto:gesteves@uepb.edu.br}{gesteves@uepb.edu.br}}, \quad
		Roberto Hirata Jr\footnote{University of S\~ao Paulo, S\~ao Paulo-SP, Brazil. email: 
			\href{mailto:hirata@ime.usp.br}{hirata@ime.usp.br}}
	} 
}

\date{November, 2015}

\definecolor{blue}{RGB}{41,5,195}

\hypersetup{
	pdftitle={\@title}, 
	pdfauthor={\@author},
	pdfsubject={Artigo científico},
	pdfcreator={LaTeX},
	pdfkeywords={latex}{atigo científico}{Esteves and Hirata Jr}, 
	colorlinks=true,       		
	linkcolor=blue,          	
	citecolor=blue,        		
	filecolor=magenta,      		
	urlcolor=blue,
	bookmarksdepth=4
}

\begin{document}

\maketitle

\thispagestyle{fancy} 

\begin{abstract}
 \noindent Microarray technology is still an important way to assess gene expression in molecular biology, mainly because it measures expression profiles for thousands of genes simultaneously, what makes this technology a good option for some studies focused on systems biology. One of its main problem is complexity of experimental procedure, presenting several sources of variability, hindering statistical modeling. So far, there is no standard protocol for generation and evaluation of microarray data. To mitigate the analysis process this paper presents an \texttt{R} package, named \texttt{maigesPack}, that helps with data organization. Besides that, it makes data  analysis process more robust, reliable and reproducible. Also, \texttt{maigesPack} aggregates several data analysis procedures reported in literature, for instance: cluster analysis, differential expression, supervised classifiers, relevance networks and functional classification of gene groups or gene networks.\\
 \textbf{Keywords:} Gene Expression. \texttt{R} Software. Statistical Methods.
\end{abstract}

\section{Introduction}
\label{pack}

Microarray technology is still an important way to assess gene expression in molecular biology. A quick Google Scholar search (31/03/2015) shows about 37,300 results mentioning this technology during 2014. The reason for this number of papers is mainly because one can measure expression profiles for thousands of genes at same time, what turns the technology a good option for some studies focused on systems biology.

One of the main problems with this technology is that the experimental procedure is complex and the data has several variability sources \citep{Draghici2006a}, that makes the statistical modeling more difficult. Besides that, the amount of numerical data and non-numerical data to deal with in a rich experiment is challenging. Beginning with the raw values given by the image analysis software, the analysis process unfolds in several steps, each one generating new and large data that are difficult to organize, keep track and later, reproduce. 

These characteristics turn data analysis a very complex and dynamic process, where most adequate data analysis procedures must be identified (or proposed) and implemented. In this sense, a computational environment that facilitates integration of several data analysis methods already proposed would be of great help. Furthermore, this environment should be implemented in a way that facilitates introduction of new algorithms, as they are proposed so frequently at literature, and also that facilitates the documentation of all methods used during data analysis process.  

To mitigate this process, this paper presents a computational package implemented into \texttt{R} system. The package is entitled \texttt{maigesPack} \citep{maigesPack}, after the acronym MAIGES that stands for Mathematical Analysis of Interacting Gene Expression Systems~\footnote{The acronym has been chosen also because maiges is also the name given to ordinary illiterated people who set themselves up as physicians.}. The main implementation integrates several data analysis packages already available in both CRAN repository and BioConductor project repository \citep{gentleman_r80_2004}. Some computational intensive methods were implemented in \texttt{C} language and can be invoked as \texttt{R} functions. Data input and also some analysis results are organized to facilitate the process. In this way, data analysis is more robust, reliable and reproducible.

The remainder of this paper is organized according to the following: Section~\ref{review} presents a Review of some available software to analyze microarray data. Section~\ref{software} presents a brief introduction about \texttt{maigesPack} organization. Section~\ref{maigesPack} shows package implementation structure. Section~\ref{examples} gives some data analysis examples with corresponding \texttt{R} code. Finally, in Section~\ref{conclusion}, main conclusion and some package future plans are presented.

\section{Review of existing software}
\label{review}

A great problem in microarray data analysis is that different computational tools use specific data structures and/or implement different mathematical and statistical methods, what difficult the interrelation between these different data analysis methods. Besides \texttt{R} briefly mentioned above, there are some programs for this kind of analysis implemented into another computational languages, like \texttt{Java}, as \texttt{MeV} \citep{saeed_374_2003} and \texttt{Cytoscape} \citep{shannon_2498_2003}.

Inside \texttt{R} software, there are several packages focused on specific steps of microarray data analysis, like \texttt{limma} \citep{smyth_2005} for linear models and \texttt{marray} \citep{yang_2006} or \texttt{OLIN} \citep{futschik_r60_2004,futschik_2006} for exploratory analysis and normalization. Thus, we developed an environment that integrates many of these packages along with another data analysis methods aimed at global microarray data analysis process. Implementation was done aiming to ensure reproducibility and process documentation capability. 

Maybe the most famous project associated to microarray data analysis is Bioconductor\footnote{\url{http://www.bioconductor.org}}, \citep{gentleman_r80_2004}, already cited above. Recently, \citet{Shen2012}
proposed a web application, known as \texttt{ArrayOU}, integrating several Bioconductor packages and \texttt{BioPerl} modules specifically designed to process Agilent microarray data. However, it is only limited to differential gene expression analysis.

Similarly to Bioconductor project, \citet{Chen2012} presents a graphical interface, implemented in \texttt{R}, to integrate some packages from the project. However, instead great contribution of the GUI interface, they also works only with differential gene expression, specially focused on Affymetrix and Illumina platforms.

In other sense, \citet{Morris2008a} proposed a suite of \texttt{Perl} modules, known as \texttt{PerlMAT}, to deal with microarray data, since image manipulation to some data analysis methods. Also, \citet{Infantino2008} presented an approach to construct a simulated microarray image, mainly focusing into evaluation of image segmentation software. In the same line, \citet{Belean2011} shows a work to compare and implement microarray image processing techniques in an automated and less computational-time sense through parallel computation.

Given the amount of public microarray datasets, nowadays it is also possible to compare various studies, taking care off eventual problems to merge datasets from different platforms, in a kind of meta-analysis of microarray data. In this way, \citet{Heider2013a} present a new \texttt{R}/Bioconductor package to do this kind of work.

Another authors have worked in new insights to use microarray data in a non-traditional analysis methods, like microarray microdissection with analysis of differences \citep{Liebner2013a}, or just using microarray technique in their specific research areas \citep{Christadore2014,Shen-Gunther2013}.

Some efforts have also been done to compare and integrate traditional mRNA microarrays with new technologies recently arised, like RNA-seq \citep{Chavan2013,Hanzelmann2013a}, microarray expression profiling specific for microRNAs \citep{Brock2013a}, or even for arrays specifically created to evaluate changes in DNA methylation, protein phosphorylation states, etc \citep{Wrzodek2013d}. Some of these new experimental approaches are based on Next Generation Sequencing (NGS) and its use is growing inside academic community \citep{Shendure2008}.

As can be seen here, there are several programs and specially \texttt{R} packages to deal with microarray data. But, some of them just implement specific methods (like gene expression analysis) or, generally, they did not integrate with each other. So, a computational environment capable to do this in an efficient and secure way is important. This was the main focus of this work, and \texttt{maigesPack} package showed here represents a step in this direction.

\section{\texttt{maigesPack} organization}
\label{software}

R is a very rich environment for scientific data organization, analysis and reporting. Its packaging system  structure makes it easy for a scientist to organize data into classes and their methods to operate on them. A packaging system is also available to mitigate the creation of new packages. It creates the directory structure for a new package and also some default files.

\begin{figure}[p!]
	\centering
	\includegraphics[width=18cm, angle=90]{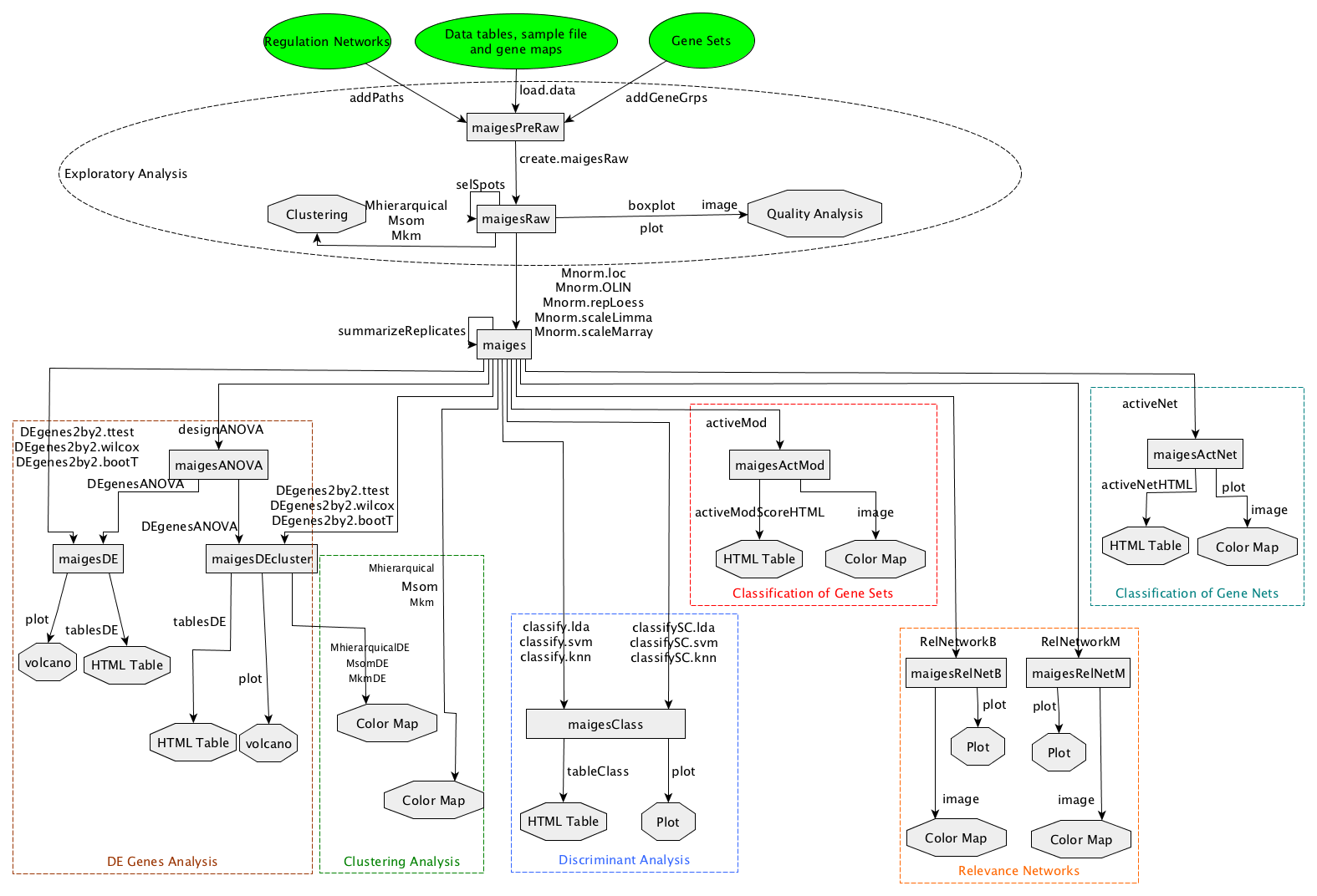}
	\caption{\label{funcoes} Classes and methods implemented in \texttt{maigesPack}. Gray rectangles represent object classes, gray octagons represent graphical or textual outputs. Edges represent mappings between objects of different types. Rectangles in dashed lines represent different data analysis modules.}
\end{figure}

The package presented in this section, \texttt{maigesPack}, is an example of a package created using this system. Figure~\ref{funcoes} presents the main object classes and methods that were implemented in this package. The figure shows an oriented and acyclic  graph where gray rectangles represent object classes, gray octagons represent graphical or textual outputs (like figures, colormaps, HTML or text files, and so on) and edges represent computational methods that act over the initial objects giving raise to new objects that they point to. The rectangular blocks in dashed lines (with different colors) present different modules grouping several specific types of data analysis methods, like differential gene expression search, classification methods, gene set expression analysis and relevance networks.

Practically, the graph present statistical and mathematical methods implemented so far that can be used to analyze microarray data. The class \texttt{"maiges"} is a central environment class, that is, all data analysis methods acts over objects from this class. Each edge represents one or more computational methods used for a specific data analysis process. It is important to notice that the use of these computational methods involves specification of several parameters, what may not be a trivial task. This design of computational classes and methods confer modularity to data analysis process, making the specification of some  parameters a simpler task, since each procedure is represented by a well defined sequence of methods in a single way. Furthermore, this feature provides an easy way to extending computational environment, since new methods can be implemented in new analysis modules.

\begin{figure}[ht!]
	\centering
	\includegraphics[width=13cm]{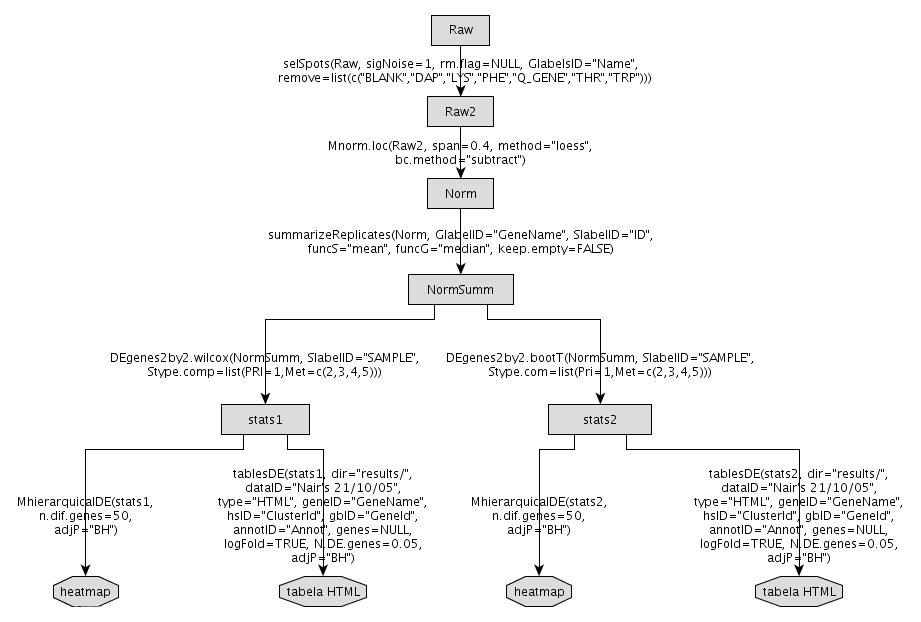}
	\caption{\label{realizacao} Instantiation graph  example. This graph   represents an specific analysis, i.e., differently from Figure~\ref{funcoes}, gray rectangles represent outputs of objects, while edges represent methods, along with used parameters.}
\end{figure}

The graph into Figure~\ref{funcoes} helps to mitigate the design of entire data analysis process that can be seen as a three-step procedure. The first one is to draw a graph of an analysis procedure, where methods to be used must be identified accordingly to statistical modeling. In the second step, computational methods already implemented should be identified for each analysis. In this step, new methods that are not yet implemented in the environment are also identified. Finally, in the third step the parameters associated with computational methods to be used should be adjusted. This last step defines an instantiation graph (see Figure~\ref{realizacao}), that is, a graph for the statistical analysis procedure. Computational methods and their parameters are recorded by this instantiation graph. Interpretation of this figure is similar to Figure~\ref{funcoes} but, in this case, edges represent methods along  with used parameters. The object \texttt{stats1} into Figure~\ref{realizacao}, for instance, represents the result of differentially expressed genes analysis, using Wilcoxon test. A key difference between both environment and instantiation graphs is that in the latter each edge of the graph represents a single analysis method that was applied to source object, generating target object. Finally, the graph can be later transcribed to a script that runs automatically in the environment.   

Figure~\ref{realizacao} presents two data analyzes aiming to find differentially expressed (DE) genes. One of them using Wilcoxon test (\texttt{DEgenes2by2.wilcox} method) and the other one using re-sampling test (\texttt{DEgenes2by2.bootT} method). In addition, were also used a method to do hierarchical clustering from objects generated by DE analysis. Notice that entire data analysis process starts from an object of class \texttt{"maigesRaw"} and so, following a graph path it is possible to follow employed methods until completion of the process, such as construction of hierarchical clustering or generation of HTML tables.

These characteristics of our computational environment give reproducibility to data analysis process because, once the instantiation graph was defined and executed, it can be stored for an entire data analysis documentation. Once the graph is stored, all data analysis can be re-executed later, without any difficult. Furthermore, if any parameter needs to be altered in any method, or if some data information needs to be modified (like slide or patient annotation), a new instantiation graph will be created in a simple and efficient way, and all data analysis can be easily re-executed, which confers robustness to the process.

\section{\texttt{maigesPack} implementation} 
\label{maigesPack}

As mentioned above, several computational programs for microarray data analysis have been proposed. Most of them are implemented into \texttt{R} that is specially interesting for this type of analysis, since it has several statistical methods and graphical tools implemented, and naturally it was used to implement \texttt{maigesPack}.

In this environment, it was used a procedure to load gene expression data that ensures data integrity regardless the format of numeric tables generated by the image processing program. Additionally, several different packages from CRAN and BioConductor specifically designed for processing, normalization and basic data analysis have been included. Tables~\ref{pacotesR} and~\ref{pacotesBioc} describes used packages both from CRAN and BioConductor, respectively. 
\begin{table}[ht!]
	\centering
	\caption{\label{pacotesR} CRAN packages used into \texttt{maigesPack}.}
	\begin{tabular}{l|l} \toprule
		\textbf{CRAN Package}            &  \textbf{Description}                                             \\ \midrule
		\texttt{amap} \citep{amap}   &  distance functions and $k$-means clustering \\
		\texttt{class} \citep{venables_2002}   &  $k$-neighborhood classification                     \\
		\texttt{e1071}  \citep{dimitriadou_2006} &  SVM classification                                            \\
		\texttt{gplots} \citep{gplots}&  graphical tools for colormaps                           \\
		\texttt{MASS} \citep{venables_2002}   &  Fisher's linear discriminant                               \\
		\texttt{R2HTML} \citep{r2html}&  write HTML pages from R objects                  \\
		\texttt{rgl} \citep{rgl}           &  tri-dimensional graphical tools                         \\
		\texttt{som} \citep{som}     &  SOM clustering                                                   \\ \bottomrule
	\end{tabular}
\end{table}

\subsection{Object classes}

In this computational environment we defined some object classes to manage since entire dataset, both raw and normalized, up to specific results from data analysis methods. Initially, we defined an object class that receives all numerical data and cases information. This class is called \texttt{"maigesPreRaw"} and it is formed by lists that stores numeric raw values obtained from data tables generated by image analysis software, text vectors specifying gene groups, graphs representing gene regulatory networks, numerical values specifying chip configuration and labels for genes and observations. It was also added into this class a logical vector to specify bad spots and another string slot that can be used to store any additional or relevant information about the study. 
\begin{table}[ht!]
	\centering
	\caption{\label{pacotesBioc} Bioconductor packages used into \texttt{maigesPack}.}
	\begin{tabular}{l|l} \toprule
		\textbf{BioC. Packages}                         &  \textbf{Description}                                              \\ \midrule
		\texttt{annotate} \citep{annotate}  &  write HTML pages with genomic annotation     \\
		\texttt{convert} \citep{convert}      &  convert between different object classes          \\
		\texttt{graph} \citep{graph}           &  graph manipulation                                           \\
		\texttt{limma}                                     &  normalization and DE analysis by ANOVA \\
		\texttt{marray}                                    &  normalization and exploratory data analysis       \\
		\texttt{multtest} \citep{multtest}  &  p-value multiple tests adjustment                      \\
		\texttt{OLIN}                                      &  normalization by OLIN and OSLIN methods         \\ \bottomrule
	\end{tabular}
\end{table}

This object class acts as an entrance hall at \texttt{R} which defines an intermediate step, previous to data analysis. In this intermediate phase it is possible to do  exploratory analyzes seeking to identify eventual problems, this can result in an update in the bad spots slot. A \texttt{"maigesPreRaw"} object is created using the function \texttt{loadData}, there are also specific functions to load information about gene groups and networks. 

From \texttt{"maigesPreRaw"} class presented before, we can create objects of \texttt{"maigesRaw"} class using a method to be presented latter. The main idea here is to store raw data after the exploratory analysis, where the main problems were already identified and data are ready for normalization step. This class defines four numerical matrices to store foreground and background intensities for both channels one and two, besides two logical matrices, one of them to index the spots to be used into normalization step and another to index spots into gene groups. Finally, three lists store gene networks and labels for samples and genes.

Another object class implemented here is named \texttt{"maiges"} that is generated from \texttt{"maigesRaw"} class using specific data normalization methods, discussed in more details latter. The matrices that stores intensity values at \texttt{"maigesRaw"} class are replaced by other two matrices here containing log-ratio intensity values (denoted here as $W$) and mean intensity values ($A$), also in logarithmic scale. Besides these two matrices, this class also defines another three ones that can store standard deviation and confidence intervals for normalized values. These values are calculated through an iterative process that repeats the normalization method selecting random spot groups. The remaining fields of the \texttt{"maigesRaw"} are kept into \texttt{"maiges"} class. Special attention is given to the class \texttt{"maigesANOVA"}, which is an extension of \texttt{"maiges"} class containing two additional slots, one of them defines a matrix for coefficients/contrasts used to fit an analysis of variance model (ANOVA) and the other one stores results from parameter estimation. 

All object classes defined in this environment were implemented in a way that some slots aren't mandatory. For instance, if the user will not do any analysis that needs gene network information, the slot associated with gene networks may stay empty in the object. In case the user try to do an analysis that require this slot, the method will prompt a warning or error message, while all other methods that do not need this information remains working without any problem. Another object classes were also implemented to store specific data analysis results, as described below: 
\begin{itemize}
	\item \texttt{"maigesDE"} - class to store results from differential gene expression analysis;
	\item \texttt{"maigesDEcluster"} - an extension to the previous class, which adds a matrix of log-ratio $W$ for a posterior cluster analysis;
	\item \texttt{"maigesClass"} - class to store results of discriminant (or classification) analysis; 
	\item \texttt{"maigesActMod"} - class to store results from functional classification of gene sets; 
	\item \texttt{"maigesRelNetB"} - class to store results from relevance network analysis \citep{butte_12182_2000};
	\item \texttt{"maigesRelNetM"} - class to store the results from an adapted method for relevance network analysis; 
	\item \texttt{"maigesActNet"} - class to store results from functional classification of gene networks. 
\end{itemize}

It is important to mention that these classes makes an efficient and robust structure for the computational environment created to microarray data analysis. Also we defined some methods for conversion between the main classes \texttt{"maigesRaw"} and \texttt{"maiges"} in other classes defined at \texttt{marray} and \texttt{limma} packages. This turned possible to use all available methods from both packages without major problems in our environment. We also created reciprocal methods, that is, methods that convert objects from \texttt{marray} and \texttt{limma} packages to the classes defined here. Below, we present briefly, data analysis methods that were done in this work. All methods acts over objects from classes presented here giving as output other objects, graphics and text or HTML files. Some examples will be given at Section~\ref{examples}.

\subsection{Data load procedure}

The procedure proposed here aims to guarantee data integrity. It is important to note that this task needs some care outside \texttt{R} environment, where all data tables must be checked for presence of any kind of problem, related with differences in line numbers, character coding, data fields with less (or more) elements than others, and so on. Our experience with microarray data analysis had showed several problems like these.

Once these evaluation were done, one may use the \texttt{loadData} method to load all data information. This method generates an object of class \texttt{"maigesPreRaw"}, and receives just one parameter specifying a configuration file. That specifies all necessary information for data load process.

Figure~\ref{conf} shows an example of a configuration file from a real dataset worked by our group. As a result from \texttt{loadData} method we have an object of \texttt{"maigesPreRaw"} class. After generation of this object, it is possible to use methods \texttt{addGeneGrps} and/or \texttt{addPaths} to load information from gene groups (like GO categories) and gene regulation networks (like KEGG pathways), respectively. 
Currently we are working to improve these functions to use methods from \texttt{GO.db}, \texttt{KEGG.db} and \texttt{KEGGgraph}, as well as looking to use other annotation packages like \texttt{org.Hs.eg.db}.
\begin{figure}[ht!]
	\centering
	\begin{tabular}{|l|} \hline
		\texttt{dataDir = "data"} \\
		\texttt{ext = NULL} \\
		\texttt{sampleFile = "sample\_file\_new\_260107.txt"} \\
		\texttt{datasetId = "Test Dataset"} \\
		\texttt{geneMap = "map48k\_07\_12\_04.txt"} \\
		\texttt{headers = c('Ch1 Mean', 'Ch1 B Mean', 'Ch2 Mean', 'Ch2 B Mean', 'Flags')} \\
		\texttt{skip = 62} \\
		\texttt{sep = ","} \\
		\texttt{gridR = 12} \\
		\texttt{dridC = 4} \\
		\texttt{printTipR = 10} \\
		\texttt{printTipC = 10} \\ \hline
	\end{tabular}
	\caption{\label{conf} An instance of a configuration file used as a parameter for \texttt{loadData} method. Each line represents an information according with items above.}
\end{figure}

Once generated the initial \texttt{"maigesPreRaw"} class object, it must be converted into a \texttt{"maigesRaw"} class object using the function \texttt{createMaigesRaw}. This method receives an instance of \texttt{"maigesPreRaw"} and parameters specifying the numerical fields that stores background and foreground intensity values for both channels that will be used for normalization and data analysis. Furthermore, this method may also use two string parameters giving gene labels that will be used to map gene groups and gene networks to respective gene labels, case gene sets and networks would be used.

\subsection{Methods for exploratory data analysis}

Previous to any kind of normalization or data analysis, a careful exploratory analysis is extremely important. This work must be done aiming to look for potential problems or pitfalls in the dataset, that may result in systematic effects. So this initial work is very important to guide correct identification of data normalization methods. As mentioned above, we did not define any specific method for exploratory data analysis into an instance of class \texttt{"maigesPreRaw"}. However, all standard \texttt{R} graphical tools, like scatter plots, boxplots, etc, may be executed directly. 

For instances of class \texttt{"maigesRaw"}, it may be interesting to do some descriptive graphics, like MA plots, boxplots and chip image representations. These can be done using some graphical tools implemented inside both \texttt{marray} and \texttt{limma} packages. In our work, we also created methods for the conversion of objects from classes \texttt{"maigesRaw"} and \texttt{"maiges"} to classes \texttt{"marrayRaw"} and \texttt{"marrayNorm"}, respectively (for package \texttt{marray}), and also to convert same classes to \texttt{"RGList"} and \texttt{"MAList"} (for package \texttt{limma}). This turned possible to use, inside our environment, all methods already implemented into both packages. For descriptive graphical tools, the methods from \texttt{marray} package are used as standard, once they are more generic and present more graphical representation options. It is also possible to convert objects between classes defined by \texttt{limma} and \texttt{marray} packages, using the package \texttt{convert}.

It is important to mention that in this work the common $M$ value, that is the log-ratio from \texttt{cy5} ($R$) over \texttt{cy3} ($G$) intensity values given by $M = \log_2(R) - \log_2(G)$, was redefined for the log-ratio between the interest sample, denoted by $I$, over reference sample, denoted by $C$, intensities (independent of who was marked with one or another dye), this ratio was renamed here for $W$, ie,
$$
W = \log_2(I) - log_2(C).
$$
This was done to prevent common misleading associated with experiments using dye-swap as experimental design.

\subsection{Data normalization methods}

For normalization procedures, there are several available methods, such as lowess non-linear regression \citep{cleveland_829_1979} (global or by blocks), OLIN/OSLIN \citep{futschik_r60_2004}, scale adjustments (global or by blocks) and VSN (variance stabilization normalization) \citep{huber_s96_2002}. All these normalization methods were already implemented in some bioconductor packages, specifically \texttt{limma}, \texttt{marray} and \texttt{OLIN} \citep{futschik_2006}. 

So, we developed the \texttt{normLoc} method based in some functions from \texttt{limma} package for location bias correction. We also developed two methods for scale adjustment (\texttt{normScaleLimma} and \texttt{normScaleMarray}) that use functions implemented into \texttt{limma} and \texttt{marray} packages, respectively. Were implemented another method, called \texttt{normOLIN} to do normalization based on {\em Optimized Local Intensity-dependent Normalization} (OLIN) proposed by \citet{futschik_r60_2004} and already implemented into \texttt{OLIN} package and available into BioConductor project. More details about these data normalization methods can be found into respective references or in \citet{yang_403_2003}.

Besides these normalization methods we also constructed an algorithm to repeat the lowess adjustment several times using a predefined proportion of dataset points randomly selected. This process turns possible to estimate standard deviation and confidence intervals for $W$ values after normalization step. It is important to note that several methods for background subtraction already implemented into \texttt{limma} package may be directly used into calls for our functions.

\subsection{Methods for differential gene expression analysis, clustering and classification} 

Maybe one of the most trivial microarray data analysis, but having great importance for biologists in terms of interpretation, is the search for differentially expressed (DE) genes, that consists at identification of individual genes showing distinct mean expression between two or more tissue types between the observed sample. More information about DE analysis can be found in \citet{dudoit_111_2002} or \citet{dudoit_2002}.

To do differential gene expression analysis, we used two statistical tests already implemented into \texttt{R} by default (into \texttt{stats} package), that is the Student's t test for two sample means comparison and its non parametric equivalent Wilcoxon test (that is similar to Mann-Whitney test). They were used into \texttt{deGenes2by2Ttest} and \texttt{deGenes2by2Wilcox}, to look for DE spots (thought to be representative for some genes) in two different biological conditions. Note that \texttt{stats} package belongs to main \texttt{R} installation, and so, it is not cited at Table~\ref{pacotesR}. Additionally to these classical tests, we also implemented another one based on re-sampling (or bootstrap) strategy, what gives a more robust option of non parametric test for DE analysis, the method was named \texttt{deGenes2by2BootT}. All methods mentioned here apply into \texttt{"maiges"} class objects.

For datasets with more complex experimental designs, like that with more than two factors or factorial designs, we used ANOVA models implemented into \texttt{limma} package in one method called \texttt{deGenesANOVA}. This method acts over \texttt{"maigesANOVA"} class objects that is obtained from \texttt{maiges} class objects using the \texttt{designANOVA} function, that constructs design and contrasts matrices to be used into ANOVA model \citep{smyth_2005}. 

We also did a method, \texttt{tablesDE}, to save an HTML (or CSV) file containing DE analysis results, and another three methods, that is \texttt{hierMde}, \texttt{kmeansMde} and \texttt{somMde}, to do cluster analysis based into the most DE genes selected by test p-values. All these methods use algorithms for multiple tests p-value adjustment available into \texttt{multtest} package, also from BioConductor project. Respectively, the methods uses hierarchical, $k$-means and {\em Self Organizing Maps} algorithms for clustering implemented in some \texttt{R} packages. 

Also, we constructed equivalent methods to do the same cluster analysis for entire dataset, without differentially gene expression selection, they are \texttt{hierM}, \texttt{kmeansM} and \texttt{somM}, respectively. These three methods may be applied into objects from classes \texttt{"maigesRaw"} or \texttt{"maiges"}, and may be used for specifics gene groups or networks. 

Similarly to cluster analysis, we also implemented some functions for classification techniques. In this case, we developed methods for Fisher linear discriminant analysis \citep{johnson_1995,mardia_1979}, Support Vector Machines \citep{burges_121_1998,mukherjee_59_1999,vapnik_1982,vapnik_1998} and $k$-neighbors \citep{dudoit_77_2002,hastie_2001,ripley_1996}. These methods may be used through \texttt{classifyLDA}, \texttt{classifySVM} and \texttt{classifyKNN}, respectively. In this type of analysis it is made an exhaustive search for groups of few genes (like pairs, trios or quartets) capable of distinguishing between different biological conditions, based only on their expression values. The implementation of these functions used several methods from \texttt{class}, \texttt{MASS} and \texttt{e1071} \texttt{R} packages. One major problem associated with the exhaustive search is its elevated computational cost, specially for huge datasets. To contour these limitation we implemented similar methods (\texttt{classifyLDAsc}, \texttt{classifySVMsc} and \texttt{classifyKNNsc}) using an heuristic algorithm known as search and choose, that first look for few single genes (like DE genes) and than makes exhaustive search in these few genes \citep{cristo_2003}.

\subsection{Methods for relevance networks and   functional classification of gene sets}

A kind of generalization of DE genes pointed into previous section is the classification of gene sets as DE that, differently of the previous one now classify groups of genes instead of individual ones as DE. This is biologically important to study genes associated with biological functions like Gene Ontology (GO) or gene networks. For these gene groups it is also possible to construct relevance networks that look for pairs of genes with alterations in association of their expression values.

In our environment we also implemented models for functional classification of gene networks, as proposed by \citet{segal_1090_2004} and for construction of relevance networks, proposed by \citet{butte_12182_2000}. This was done into \texttt{activeMod} and \texttt{relNetworkB} functions, respectively, using some packages from base \texttt{R} installation (specially \texttt{stats}). Beyond these methods, we also created some tools for visualization, including heatmaps to show activation/inactivation profiles of gene groups and graphical representation for relevance networks. 

Furthermore, we adapted Butte's original method for relevance networks, at \texttt{relNetworkM} function, were we look for pairs of genes that shows significant coefficient correlation alterations between two distinct biological conditions. 

Relevance networks construction requires the use of some association measure and the main coefficient used for this is Pearson's correlation, as proposed originally by \citet{butte_12182_2000}. But this coefficient can't detect non-linear associations so, to circumvent this problem, \citet{butte_415_2000} proposed the use of mutual information as association measure to construct relevance networks. This association measure was also implemented into our environment, as the \texttt{MI} method, using an algorithm proposed by \citep{kraskov_2004}. 

Another frequent problem associated with Pearson's correlation is the high susceptibility to outliers presence into dataset, what can make the resulting coefficient measure highly biased. In this way, to minimize the problem we created a more robust method, implemented as \texttt{robustCor}, that remove one extreme value using an idea similar to leave-one-out from classification techniques. 

Other important contribution of our work was the proposition of a statistical model to do functional classification of gene regulation networks, were we use information from the p-value of a zero correlation test to construct a test statistic with known probability distribution. The manuscript of this method is under production. This statistical model were implemented into \texttt{activeNet} \texttt{R} method. 

\subsection{\texttt{C} implementation of some functions}

The \texttt{R} environment is an extremely powerful tool, as it is designed as statistical computation language for data analysis and manipulation. Also, its graphical tools offers a multitude of possibilities of data visualization and presentation. However, as \texttt{R} is an interpreted language, it pays off a high computational cost for large amounts of data (what is very common in gene expression datasets), specially for methods that needs many loops and nested conditional expressions. This happens into our codes mainly for resampling of t test and calculation of robust correlation coefficient, what makes a pure  \texttt{R} implementation of these codes inefficient. 

However, it is possible to use pre-compiled code into another computational language (like \texttt{C}, \texttt{C++} or \texttt{Fortran}) inside \texttt{R} functions. As these compiled code are much faster than interpreted one, we gain a lot of time using this strategy. Much of the main \texttt{R} methods are implemented into another language (mainly \texttt{C}, \texttt{C++} or \texttt{Fortran}) and the compiled library are loaded automatically at \texttt{R} initialization. So, to optimize our functions, we implemented the most computational costly piece of code for \texttt{deGenes2by2BootT}, \texttt{robustCor} and \texttt{MI} into \texttt{C}. This gave a significant gain in computational time for these functions.

\section{Some examples}
\label{examples}

In this section we demonstrate the functionality of \texttt{maigesPack} using a dataset with some observations of gastric-esophageal data. This dataset was analyzed by our group and resulted in a research article in {\em Cancer Research} \citep{gomes_7127_2005}. Following we will show some data analysis examples using a piece of this dataset, that is included as part of the package, as \texttt{gastro} dataset. The original data have 4800 spots (approximately 4400 unique genes) and 172 chips with dye swap (86 unique samples). The sub-dataset included here has 500 spots (representing 486 unique genes) and 40 chips (20 observations). 

The implementation we did use some variables (named labels) for samples and genes. For \texttt{gastro} dataset we have three important labels for samples (or patients) and 5 important labels for genes. For patients we have \textit{Sample} that shows an unique identification to each observation, \textit{Tissue} that shows the tissue classification for each observation (\textit{Aeso} is Adenocarcinoma from esophagus, \textit{Aest} is adeno from stomach, \textit{Neso} is normal esophagus and \textit{Nest} is normal stomach) and \textit{Type} that shows the general type of the observations (\textit{Col} is columnar tissue and \textit{Sq} is squamous tissue). For genes the important labels are \textit{Name}, \textit{GeneId}, \textit{GeneName}, \textit{ClusterId} and \textit{Annot}. 

To load this dataset, start \texttt{R} in any working directory, load \texttt{maigesPack} and \texttt{gastro} dataset using the following commands: 
\begin{verbatim}
	> library(maigesPack)
	> data(gastro)
\end{verbatim}

The command \texttt{data(gastro)} above, loads five objects into \texttt{R} session. Table~\ref{objects} shows names and classes from these objects. 
\begin{table}[ht!]
	\centering
	\caption{\label{objects} Objects from \texttt{gastro} dataset, included with \texttt{maigesPack}.}
	\begin{tabular}{c|c} \hline
		\textbf{Object Name}  & \textbf{Class Object}  \\ \hline
		\texttt{gastro}      & \texttt{"maigesPreRaw"} \\
		\texttt{gastro.raw}  & \texttt{"maigesRaw"}    \\
		\texttt{gastro.raw2} & \texttt{"maigesRaw"}    \\
		\texttt{gastro.norm} & \texttt{"maiges"}       \\
		\texttt{gastro.summ} & \texttt{"maiges"}       \\ \hline
	\end{tabular}
\end{table}

So, to see all sample labels available into \texttt{gastro}, type \texttt{names(gastro@Slabels)}, analogously, type \texttt{names(gastro@Glabels)} to see all gene labels. These commands also works for the other four objects available into \texttt{gastro} dataset. 

The object \texttt{gastro.raw}, that is of class \texttt{"maigesRaw"}, is already available into dataset. The command used to generate this object is given below. 
\begin{verbatim}
	> gastro.raw = createMaigesRaw(gastro, greenDataField="Ch1.Mean",
	+   greenBackDataField="Ch1.B.Mean", redDataField="Ch2.Mean",
	+   redBackDataField="Ch2.B.Mean", flagDataField="Flags",
	+   gLabelGrp="GeneName", gLabelPath="GeneName")
\end{verbatim}

\subsection{Exploratory analysis}

\begin{figure}[ht!]
	\centering
	\begin{tabular}{cc}
		\includegraphics[width=6.5cm]{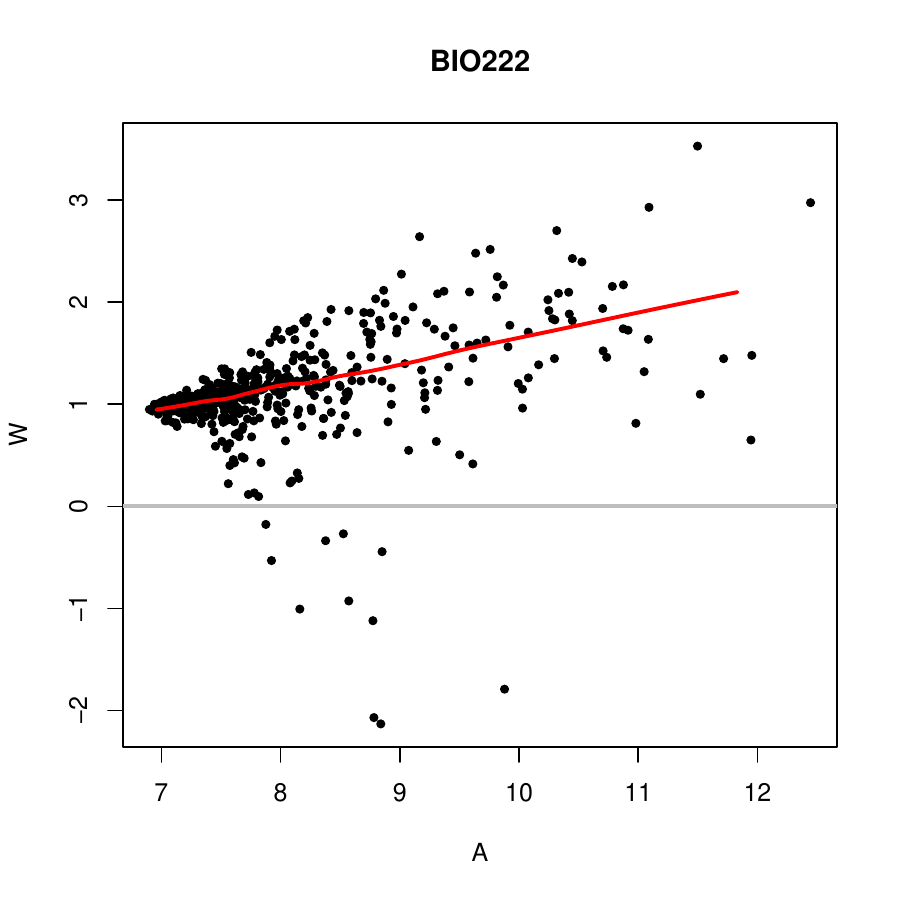}
		& \includegraphics[width=6.5cm]{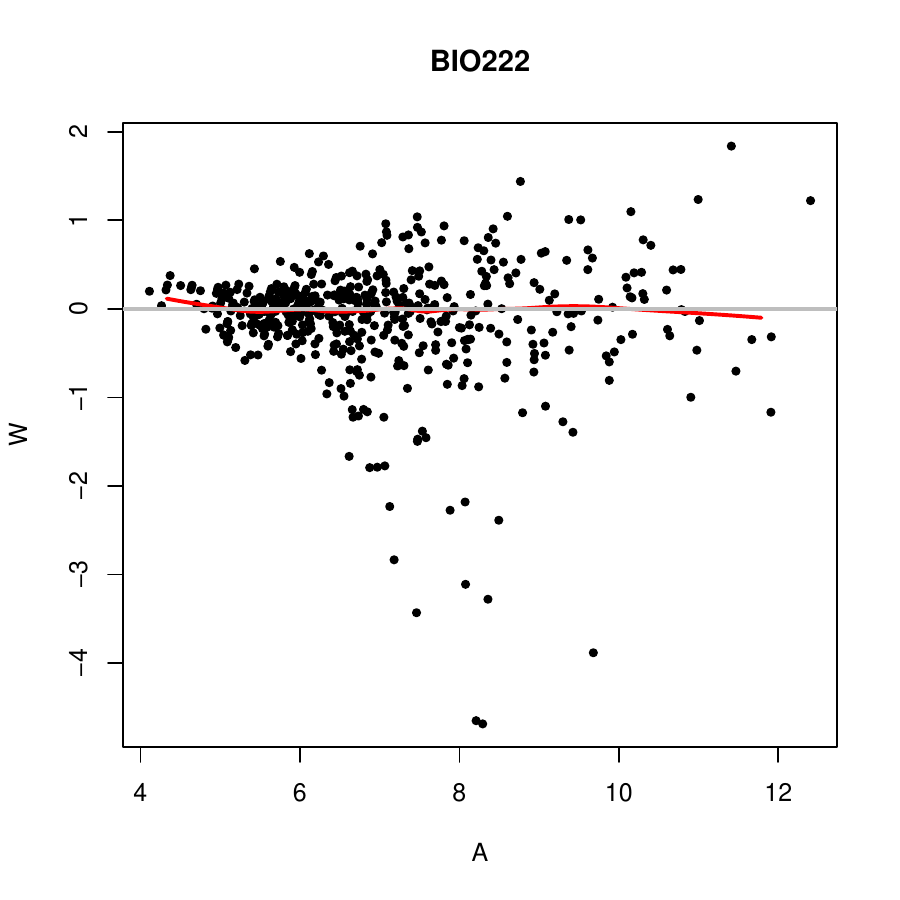} \\ 
		(a)                      &
		(b)                      \\ 
	\end{tabular}
	\caption{\label{WAplot} WA plot for the first chip of the \texttt{gastro.raw} object (a) and for \texttt{gastro.norm} object (b).} 
\end{figure}

From the object \texttt{gastro.raw} it is possible to do several exploratory analysis. Generic function \texttt{plot} can be used to show WA plots. Below there is a command to do this plot for the first chip and the result is represented at Figure~\ref{WAplot} (a). A similar plot can be done for normalized object, see Figure~\ref{WAplot} (b), but command is not represented here.
\begin{verbatim}
	> plot(gastro.raw[,1], bkgSub="none")
\end{verbatim}

\begin{figure}[t]
	\centering
	\includegraphics[width=9cm]{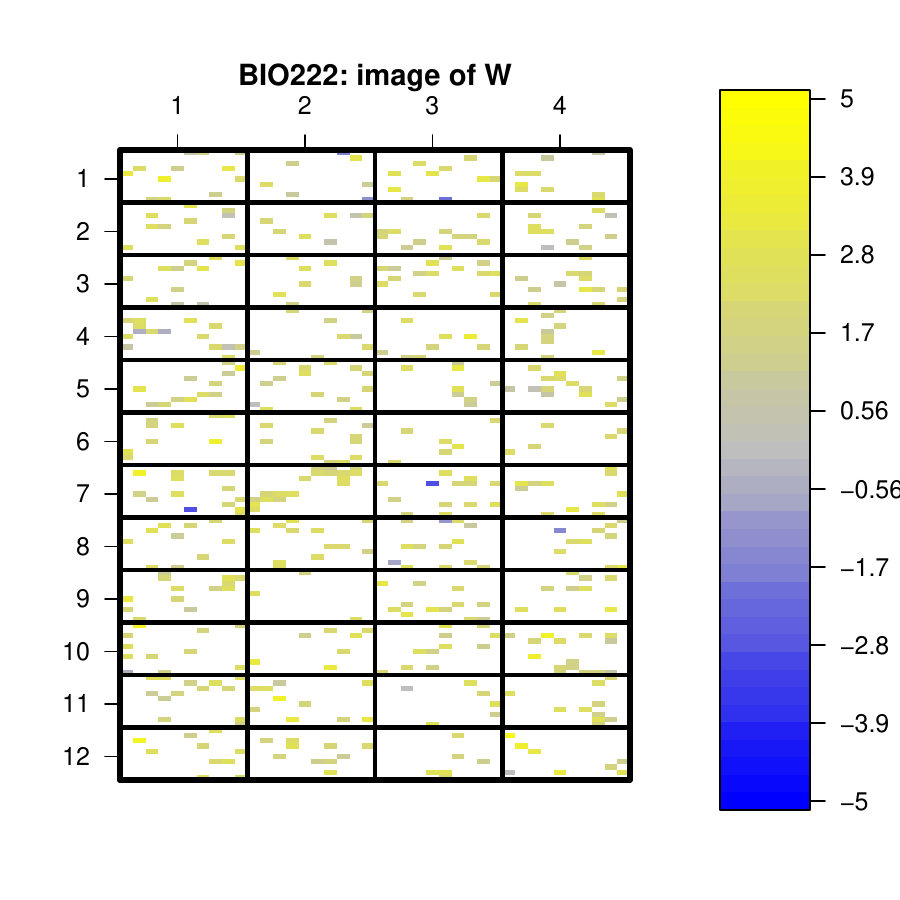}
	\caption{\label{imageW} Image of $W$ values for first chip in gastric data.}
\end{figure}

In above command, changing the parameter \texttt{bkgSub}, the user change background subtraction method, for example, if the user specify \texttt{bkgSub="subtract"} the conventional background subtraction is done. You may also do representations for numerical values into the chip using \texttt{image} method. See a command below, that represents $W$ values (the $W$ values are represented by default, that is, without any additional parameter, if additional parameters are added, conventional $M$ values are used) for the first chip that generated the Figure~\ref{imageW}, note that only some points are represented, because our example dataset has only 500 spots. 
\begin{verbatim}
	> image(gastro.raw[,1])
\end{verbatim}

To change $W$ values to another one, like $A$ for example, use \texttt{image(gastro.raw[,1], "maA")}. Take a look at \texttt{maImage} method from package \texttt{marray} to see additional parameters for this function. Another type of exploratory graphics is boxplot, that can be made using the following command, figure not shown. 
\begin{verbatim}
	> boxplot(gastro.raw[,2])
	> boxplot(gastro.raw)
\end{verbatim}

\begin{figure}[t]
	\centering
	\begin{tabular}{cc}
		\includegraphics[width=6.5cm]{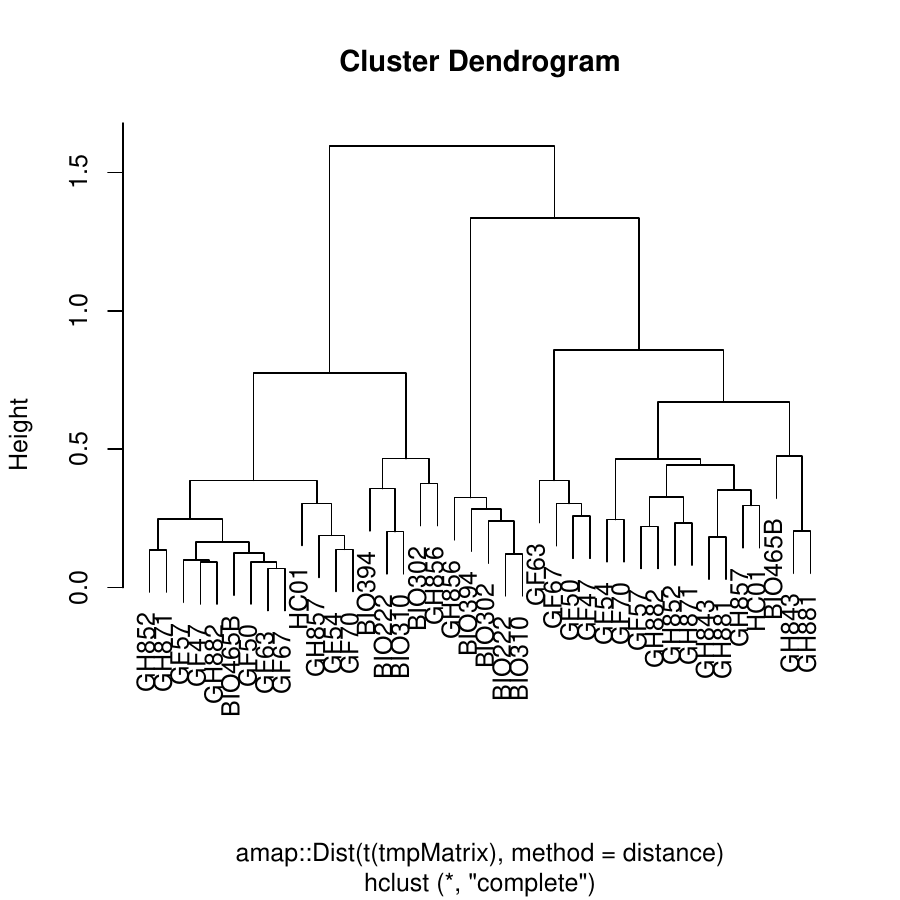}
		& \includegraphics[width=6.5cm]{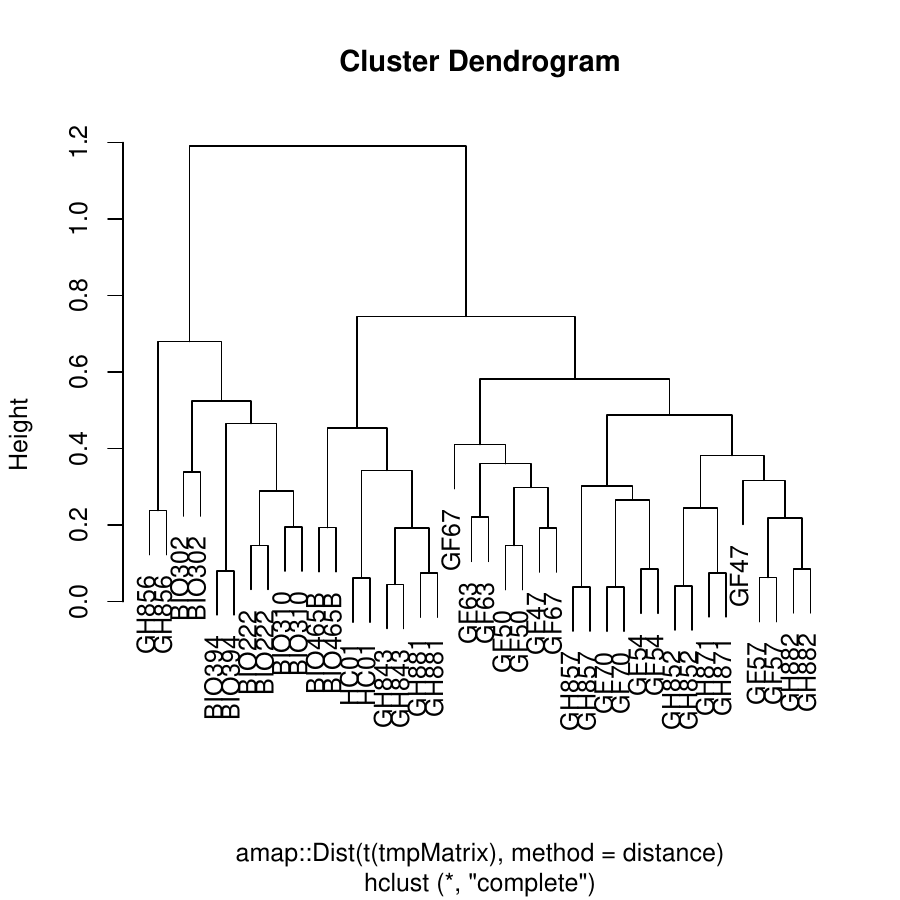} \\ 
		(a)                      &
		(b)                        \\ 
	\end{tabular}
	\caption{\label{hier} Hierarchical cluster for all observations from raw (a) an normalized (b) \texttt{gastro} dataset .} 
\end{figure}

It is also possible to do hierarchical clustering to visualize and analyze data quality. To do so the command below could be used to construct the hierarchical dendrogram presented at Figure~\ref{hier} (a).
\begin{verbatim}
	> hierM(gastro.raw, rmGenes=c("BLANK","DAP","LYS","PHE","Q_GENE","THR","TRP"),
	+   sLabelID="Sample", gLabelID="Name", doHeat=FALSE) 
\end{verbatim}

\subsection{Normalizing the dataset}

The first step to data normalization is to select spots to be used for estimating normalization factor. In this package this is done by \texttt{selSpots} function. In our example use the following command. 
\begin{verbatim}
	> gastro.raw2 = selSpots(gastro.raw, sigNoise=1, rmFlag=NULL, gLabelsID="Name",
	+   remove=list(c("BLANK","DAP","LYS","PHE","Q_GENE","THR","TRP")))
\end{verbatim}

This function automatically set spots independently for each chip (column) that will be used according with specified criteria. For example, if the user want to remove spots with signal to noise ratio less than 1.5, specify this value as \textit{sigNoise=1.5}; to remove spots marked with flags 1 and 4, specify \textit{rmFlag=c(1,4)}, and so on. 

Once this is done, the main function for normalization step is \texttt{normLoc} that uses another function from \texttt{limma} package. The most usual normalization method may be applied by the function given below. 
\begin{verbatim}
	> gastro.norm = normLoc(gastro.raw2, span=0.4, method="loess")
\end{verbatim}

There are other useful methods for data normalization, like OLIN method available in an eponymous package, and another method that repeats lowess fitting calculating standard variation and confidence interval for $W$ values. These methods can be used with commands below, but they must be used with caution because both of them are computational intensive, so it is interesting to check their arguments. 
\begin{verbatim}
	> gastro.norm = normOLIN(gastro.raw2)
	> gastro.norm = normRepLoess(gastro.raw2)
\end{verbatim}

After locale normalization, it is possible to use functions like \texttt{normScaleMarray} or \texttt{normScaleLimma} to do scale adjustment. The first method use a function from \texttt{marray} package and the second use a function from \texttt{limma} package. To do scale adjustment between print tips use the command below. 
\begin{verbatim}
	> gastro.norm = normScaleMarray(gastro.norm, norm="printTipMAD")
\end{verbatim}

To adjust the scale between chips estimating the adjustment factor by MAD use the following command.  
\begin{verbatim}
	> gastro.norm = normScaleMarray(gastro.norm, norm="globalMAD")
\end{verbatim}

In the other hand, \texttt{limma} package offers other possibilities to scale adjustment, that were also incorporated into \texttt{maigesPack}. For example, to do scale adjustment by chips using an estimator slightly different from MAD, use: 
\begin{verbatim}
	> gastro.norm = normScaleLimma(gastro.norm, method="scale")
\end{verbatim}

Another possibility is to do variance stabilization along $A$ values for all chips. But this method must be applied directly over the raw object. This may be done using the command below. Again, pay attention with this method, because it is also very time consuming. 
\begin{verbatim}
	> gastro.norm = normScaleLimma(gastro.raw2, method="vsn")
\end{verbatim}

All normalization functions generate objects of class \texttt{"maiges"}. The exploratory functions exemplified for raw objects may also be used with this class, in same way they are applied into that objects Figure~\ref{WAplot} (b) presents the normalized version of WA plot for the first chip. Also the hierarchical clustering must be done after normalization step. So in this dataset (with dye swaps), we can observe the replicate pairing for most pacients  in dye swap. This can be seen at Figure~\ref{hier} (b), compare the swap pairings into (a) and (b) figures. 

Sometimes, the sequence (or genes) fixed onto spots may be replicated. The same thing may happen with the observations (the most common is dye swap). So depending on the statistical models used, the user must have to resume data both for spots and biological observations. These may be done by using the function \texttt{summarizeReplicates}, as exemplified below. If the user needs to resume only spots or samples, simply specify \textit{funcS=NULL} or \textit{funcG=NULL}. If both of them are \textit{NULL} no summary are done. 
\begin{verbatim}
	> gastro.summ = summarizeReplicates(gastro.norm, gLabelID="GeneName",
	+   sLabelID="Sample", funcS="mean", funcG="median",
	+   keepEmpty=FALSE, rmBad=FALSE)
\end{verbatim}

\subsection{Some data analysis methods}  

After pre-processing and subsequent data normalization steps, the user may use several statistical methods for data analysis. As mentioned above, \texttt{maigesPack} integrates some methods already available into \texttt{R} and/or BioConductor project. Between these methods we have cluster algorithms, differential gene expression and discrimination techniques, functional classification of gene groups or gene networks and construction of relevance networks.

\subsubsection{Differentially expressed genes} 

As mentioned into Section~\ref{maigesPack}, we implemented three main methods to search by DE genes: \texttt{deGenes\-2by2Ttest}, \texttt{deGenes2by2Wilcox} and \texttt{deGenes2by2BootT}. The first one supposes that data are normally distributed and uses t statistic to do calculations. The other two are non-parametric, avoiding data normality assumption. The use of these functions is illustrated by the following command (by t test). The other two are used in the same way, and will not be presented here. 
\begin{verbatim}
	> gastro.ttest = deGenes2by2Ttest(gastro.summ, sLabelID="Type")
\end{verbatim}

Once the analysis was done, it is possible to see volcano plots and save HTML (or CSV) tables for the results using the commands below, respectively. 
\begin{verbatim}
	> plot(gastro.ttest)
	> tablesDE(gastro.ttest)
\end{verbatim}

\begin{figure}[ht!]
	\centering
	\includegraphics[width=11cm]{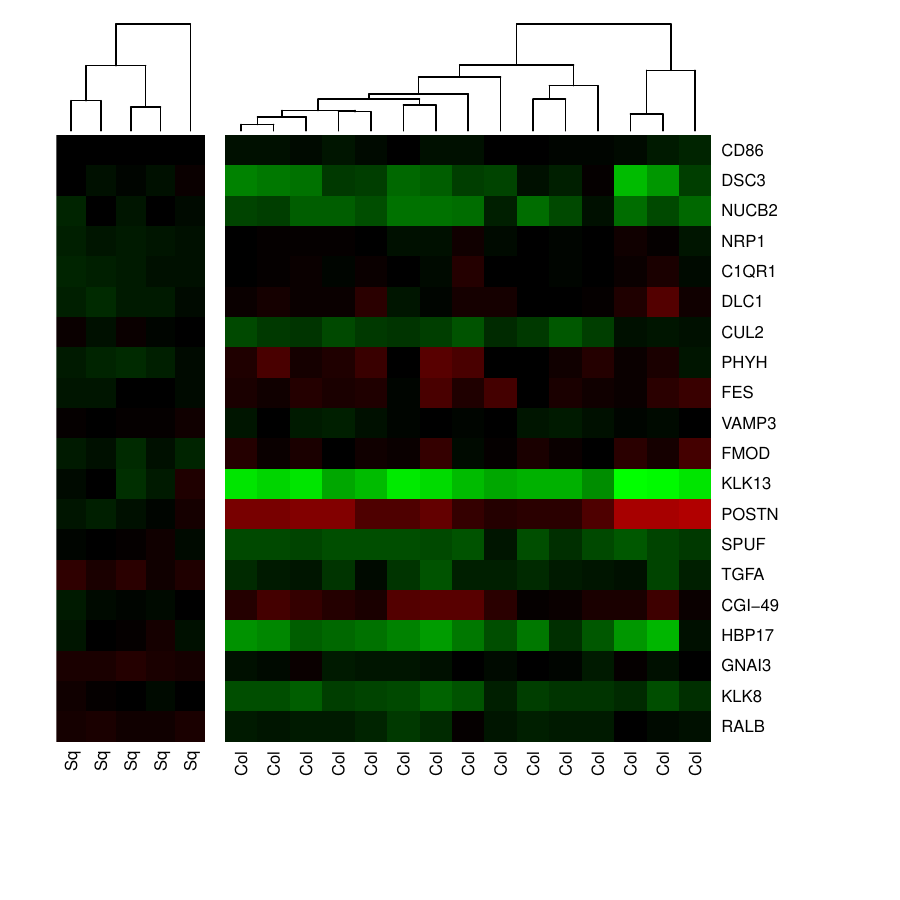}
	\caption{\label{SOM20} SOM cluster with 2 groups using 20 most DE genes.}
\end{figure}

It is also possible to do cluster analysis selecting DE genes according with tests p-values. Functions \texttt{hierMde}, \texttt{somMde} and \texttt{kmeansMde} do cluster analysis using hierarchical, SOM and k-means algorithms, respectively. To do SOM cluster with 2 groups using 20 most differentially expressed genes (adjusting p-values by FDR) use the command below, result is in Figure~\ref{SOM20}. Similarly, it is possible to do cluster analysis directly for objects of class \texttt{"maiges"}, using the functions \texttt{hierM}, \texttt{somM} and \texttt{kmeansM}, but without selecting differentially expressed genes. 
\begin{verbatim}
	> somMde(gastro.ttest, sLabelID="Type", adjP="BH", nDEgenes=20,
	+        xdim=2, ydim=1, topol="rect")
\end{verbatim}

All methods presented above are applicable to compare only two distinct biological sample types. When there are more than two types, it is possible to use ANOVA models. We also implemented one method for adjust ANOVA models, using functions from \texttt{limma}. To use this model, another method is necessary to construct design and contrasts matrices. As an example, to construct these matrices for {\em Tissue} factor (note that this factor has 4 sample types, type \texttt{getLabels(gastro.summ, "Type")} to see them), use the command below. 
\begin{verbatim}
	> gastro.ANOVA = designANOVA(gastro.summ, factors="Tissue")
\end{verbatim}

And them, to model fit and to do F test calculation use the following command. 
\begin{verbatim}
	> gastro.ANOVAfit = deGenesANOVA(gastro.ANOVA, retF=TRUE)
\end{verbatim}

If user want to do individual t tests it is possible to use (this is the default) the command below. 
\begin{verbatim}
	> gastro.ANOVAfit = deGenesANOVA(gastro.ANOVA, retF=FALSE)
\end{verbatim}

\begin{figure}[ht!]
	\centering
	\includegraphics[width=9cm]{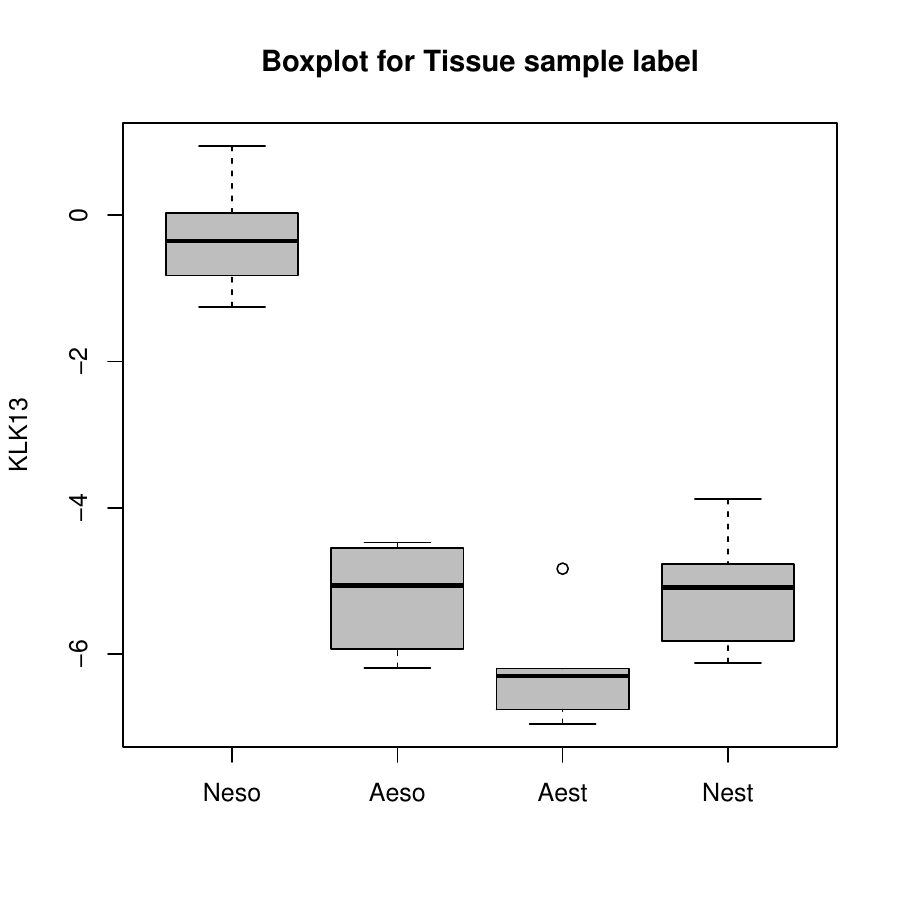}
	\caption{\label{boxplot} Boxplot for KLK13 gene separating by tissue types.} 
\end{figure}

Another function that may be useful, specially for ANOVA analysis is \texttt{boxplot}. This type of plot may help in identify tissues where some gene presented alteration. For example the gene KLK13, presented significant alteration in the F test done above. Using the command below it is possible to see in what tissue this gene is altered. The result is illustrated in Figure~\ref{boxplot}. \texttt{Boxplot} also work with \texttt{"maiges"} class objects. 
\begin{verbatim}
	> boxplot(gastro.ANOVAfit, name="KLK13", gLabelID="GeneName", 
	+ sLabelID="Tissue")
\end{verbatim}

The object resulting from this method are also of class \texttt{"maigesDE"} (or \texttt{"maigesDEcluster"}) so, commands for plot, save tables  and cluster analysis, presented above also work here. Pay attention that functions to do cluster analysis only work for objects of \texttt{maigesDEcluster} classes.

\subsubsection{Discrimination analysis}

Discrimination (or classification) analysis, that search by groups of few genes that can distinguish between different sample types was implemented in our package into \texttt{classifyLDA}, \texttt{classifySVM} and \texttt{classifyKNN} functions, as mentioned before. To evaluate grouping quality it is possible to use cross validation method.

It is also possible to search exhaustively into all genes available in the dataset, but this is very time consuming. To overcome this limitation, we added the possibility to search for classifiers exhaustively inside gene groups or networks. To search by trios of genes that classify two types (\textit{Col} and \textit{Sq}, for example) using Fisher's method, for the sixth (cell cycle arrest, gene ontology code GO0007050) gene group (into \texttt{GeneGrps} slot from \texttt{gastro.summ} object) use the following command. 
\begin{verbatim}
	> gastro.class = classifyLDA(gastro.summ, sLabelID="Type",
	+   gNameID="GeneName", nGenes=3, geneGrp=6)
\end{verbatim}

To visualize or save tables (HTML or CSV) for the resulting classifiers it is possible to use both commands below. But pay attention that \texttt{plot} only works for pairs or trios of classifiers. For trios, it will do an interactive 3-dimensional plot using \texttt{rgl} package. 
\begin{verbatim}
	> plot(gastro.class, idx=1)
	> tableClass(gastro.class)
\end{verbatim}

As mentioned before, exhaustive search is very time consuming for large gene sets and so we implemented the search and choose empiric method (\texttt{classifyLDAsc}, \texttt{classifySVMsc} and \texttt{classifyKNNsc}), that can be used the same way as exhaustive search.

\subsubsection{Functional classification of gene groups} 

Functional classification of gene groups (or modules), proposed by \citet{segal_1090_2004} basically searches for gene groups that present number of differentially expressed members greater than the expected by chance. This can be done for all gene groups loaded into the dataset (this information is stored in \texttt{GeneGrps} slot from objects of classes \texttt{"maigesPreRaw"}, \texttt{"maigesRaw"} or \texttt{"maiges"}), for instance, for sample label \textit{Tissue} using the command below. 
\begin{verbatim}
	> gastro.mod = activeMod(gastro.summ, sLabelID="Tissue", cutExp=1,
	+   cutPhiper=0.05)
\end{verbatim}

To plot results, as an image (like a heatmap), for individual observations it is possible to use the following command. 
\begin{verbatim}
	> plot(gastro.mod, "S", margins=c(15,3))
\end{verbatim}

\begin{figure}[ht!]
	\centering
	\includegraphics[width=8cm]{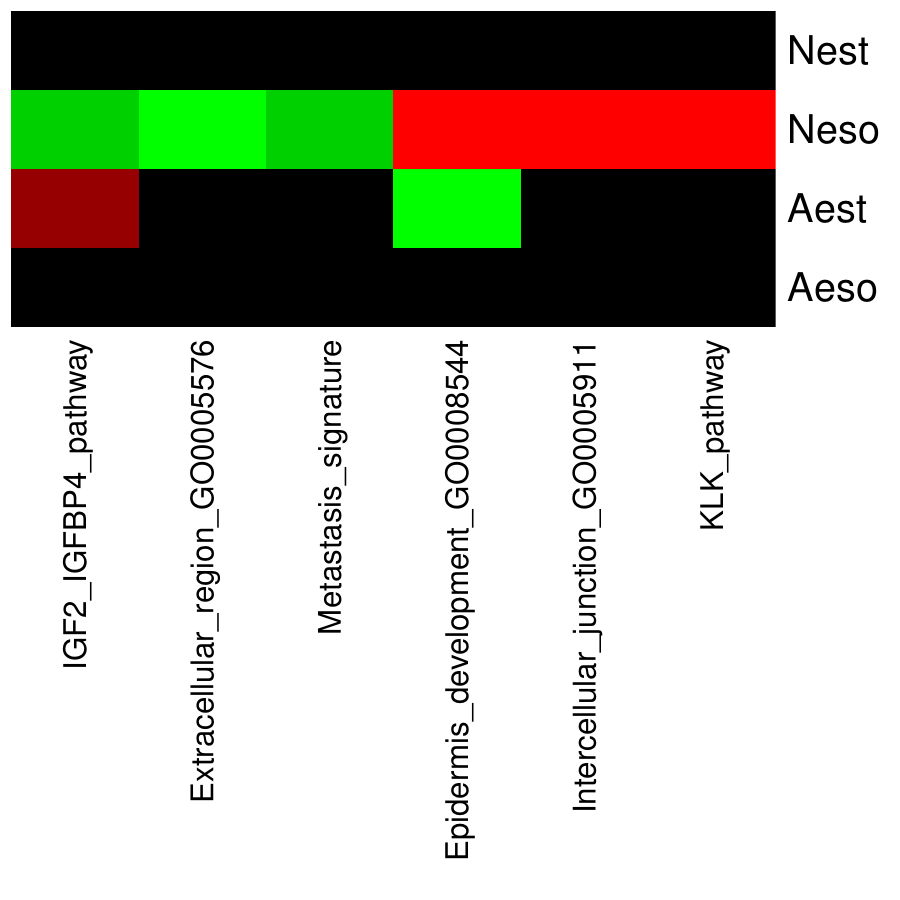}
	\caption{\label{modulo} Result from functional classification of gene groups for \textit{Tissue} label.} 
\end{figure}

The same type of graphic can be done for tissue types (or biological conditions) changing second argument from \textit{S} to \textit{C}, as below. The result is represented in Figure~\ref{modulo}. 
\begin{verbatim}
	> plot(gastro.mod, "C", margins=c(23,5))
\end{verbatim}

This method also calculate an score (with significance levels, p-values) given concordance of the gene and group classification. These values can be saved, as HTML tables, for each sample type using the command \texttt{activeModScoreHTML}.

\subsubsection{Relevance networks}

Finally, to conclude this section, relevance networks proposed by \citet{butte_12182_2000} and also discussed earlier, estimate linear Pearson's correlation values between all pairs of genes from a given gene group and test the significance of this value under the hypothesis of null correlation. Then, we select pairs of genes with correlation values significantly different from zero using some p-value cutoff. To construct relevance networks for normal esophagus (\textit{Neso}) in first gene group, for instance, use the command. 
\begin{verbatim}
	> gastro.net = relNetworkB(gastro.summ, sLabelID="Tissue", 
	+   samples="Neso", geneGrp=1)
\end{verbatim}

To visualize the results it is possible to do an image of the correlation values for all pairs of genes or a graph representing that ones with p-values less than \texttt{cutPval}. This can be done using one of the commands below. 
\begin{verbatim}
	> image(gastro.net)
	> plot(gastro.net, cutPval=0.05)
\end{verbatim}

However, Butte's original method do not make possible to compare quantitatively results obtained in two different biological conditions. So, we adapted his method to construct relevance networks with pairs of genes presenting altered correlation values between two sample types. This is done by a Fisher's $Z$ transformation. To find pairs of genes that present altered correlation values between normal tissues from esophagus and adenocarcinoma (for the seventh gene group), use the following command.
\begin{verbatim}
	> gastro.net2 = relNetworkM(gastro.summ, sLabelID="Tissue", 
	+   samples = list(Neso="Neso", Aeso="Aeso"), geneGrp=7)
\end{verbatim}

Also, it is possible to use the commands \texttt{image} and \texttt{plot} to visualize the results, as exemplified by both commands below. 
\begin{verbatim}
	> image(gastro.net2)
	> plot(gastro.net2, cutPval=0.05)
\end{verbatim}

Another graphical output from this kind of analysis is the comparison between regression lines into expression values for an altered gene pair. This can be done as showed in the next command, where we compare adjusted linear fit for genes KLK13 and EVPL both into \textit{Neso} and \textit{Aeso} tissue types.
\begin{verbatim}
	> plotGenePair(gastro.net2, "KLK13", "EVPL")
\end{verbatim}

Next section presents the main concluding remarks about \texttt{maigesPack} as well as the package availability and major next steps of our work  this package.

\section{Conclusions}
\label{conclusion}

As it was showed above, we created a computational environment based into \texttt{R} software, that integrate several mathematical and statistical tools already implemented into another \texttt{R} packages, both from CRAN and from BioConductor, as well as together another methods implemented or adapted in this work. \texttt{R} software was used because it is a statistical tool (and also a computational language) with several statistical methods, and mainly several probabilistic models already available. Also it has a great community all over the world developing tools in it, as the BioConductor project itself. This project focuses into developing statistical tools devoted mainly for genomic data analysis, as gene expression analysis, what was the focus of our work. Besides all these benefits, \texttt{R} is also free software, based into GNU public license. 

The environment presented here is intended as a modular computational framework, were new statistical and/or mathematical tools can be easily added. So, as can be observed into Figure~\ref{funcoes}, dotted rectangles represents modules for DE genes analysis, clustering, classification, construction of relevance networks, and functional classification of gene sets and networks. So, as new methods are proposed in the literature they can be added to these modular data analysis scheme. Also, new modules for another type of analysis can be easily included into this structure. 

Obviously, the computational environment presented here needs several adjustments in the implementations of several functions, since a data structure organization to several optimization in computational timings, that is our main working by this time. But, even so, it works well for an organized and fully reproducible gene expression (mainly based onto microarray technology) data analysis. Another future work that deserves attention is the adaptation of the package to deal with the technologies for large scale gene expression measurements, like RNA-seq or microRNA microarrays, that are receiving growing attention in the literature nowadays.

The \texttt{maigesPack} package that resulted from this work was submitted an approved into BioConductor project and is available from project repositories into the link
\url{http://www.bioconductor.org/packages/release/bioc/html/maigesPack.html}. The actual version of the package is 1.34.0.

\section*{Acknowledgments}
This work was financed mainly by CAPES and also by CNPq (grant 478184/2012-3). Also, we would like to thanks Raydonal Ospina Martinez and Diana Maia, for additional support to write the manuscript, and E. Jord\~ao Neves for several help and suggestions in this work.

\bibliography{esteves-hirata}

\end{document}